\documentclass[final]{aipproc}
\layoutstyle{6x9}

\usepackage{amsfonts}
\usepackage{amsmath}
\usepackage{amssymb}

\newcommand{\be}{\begin{equation}}
\newcommand{\ee}{\end{equation}}
\newcommand{\ba}{\begin{eqnarray}}
\newcommand{\ea}{\end{eqnarray}}

\begin{document}

\title{The status of the KSS bound and its possible violations
(How  perfect can a fluid be?)}

\classification{11.15.Pg, 12.38.Mh, 25.75.q, 51.20.+d}
\keywords {Viscosity over entropy density, KSS holographic conjecture, 
heavy ion collisions}

\author{Antonio Dobado\footnote{Invited speaker, contact 
{\tt{dobado@fis.ucm.es}}} }      {
address={Depto. F\'isica Te\'orica I, Universidad Complutense de 
Madrid, 28040 Madrid, Spain} }
\author{Felipe J. Llanes-Estrada}{}
\author{Juan Miguel Torres Rinc\'on}{}

\begin{abstract}
In this work we briefly review the  Kovtun-Son-Starinet (KSS)
computation of the ratio $\eta/s$ for quantum field theories with
gravitational dual and the related conjecture that it is bound from below
by $1/4\pi$. We discuss the
validity of the bound and the nature of its possible violations,
its relevance for RHIC, its connection  with phase transitions
and other related issues.
\end{abstract}

\maketitle

\section{Introduction}
Probably one of the most astonishing and relevant discoveries of
the last decade in theoretical physics was the AdS/CFT (Anti de
Sitter/Conformal Field Theory) correspondence \cite{AdS}. This
powerful tool makes it possible to get deep insights into the
dynamics of strongly interacting gauge theories with a gravity dual
in the limit $N_c \rightarrow \infty$ and $\lambda \rightarrow
\infty$, where $N_c$ is the number of colors and $\lambda$ is the
't Hooft coupling $g^2N_c$. According to this holographic
correspondence it is possible to relate some black-brane
configurations in higher dimensions to certain four dimensional
conformal Quantum Field Theories (CFT) at finite temperature. The
thermodynamic properties such as temperature and entropy of the
CFT can thus be obtained from the corresponding Bekenstein
definitions for the black-brane. Moreover, not only the thermodynamics
but also the hydrodynamic behavior of the black-brane horizon can
be identified with the CFT hydrodynamics. In particular, by using
the AdS/CFT correspondence it has been possible to compute, for a
very large class of four dimensional theories with gravity duals,
the  ratio \cite{son1}:
\begin{equation} \label{thebound}
\frac{\eta}{s}=\frac{1}{4\pi}
\end{equation}
where $\eta$ is the shear viscosity and $s$ is the entropy
density, dimensionally both scaling as $[E^3]$ in natural units. 
From the CFT point of view, the standard relativistic
hydrodynamic equations \cite{landau}, 
$\partial_\mu T^{\mu\nu} = 0$
for $T^{\mu\nu}$, the energy momentum tensor, can be
understood as the effective theory describing the dynamics of
the CFT at large distance and time \cite{son2}. Following the
philosophy of the effective theories it is possible to expand the
energy-momentum tensor in powers of the space-time derivatives. To
zeroth order we get the well known ideal fluid equations.
Dissipative processes require going to the next order where
diffusion coefficients such as  the  shear viscosity $\eta$, the
bulk viscosity $\zeta$ and the heat conductivity $\kappa$ appear
together with other possible diffusion coefficients $D_i$. From
the linearized equation of motion in momentum space at this order
one can get the dispersion relation of the  mechanical
excitations, which are two transverse modes with
\begin{equation}
\omega(k)=-i\frac{\eta}{\epsilon+P}k^2
\end{equation}
(where $\epsilon$ is the energy density and $P$ the pressure) and
one sound mode:
\begin{equation}
\omega(k)=c_s-\frac{i}{2}(\frac{4}{3}\eta+\zeta)\frac{k^2}{\epsilon+P}
\end{equation}
where $c_s=\sqrt{dP/d\epsilon}$ is the speed of sound. 
The imaginary part in the right hand-side entails a damping
of the corresponding mode, and its size therefore measures the ability
of the fluid to dissipate perturbations.
Since for vanishing chemical potential we have the thermodynamic equation
$\epsilon + P= Ts$, $\eta/s$ governs the right hand side and
is a good way to characterize
the intrinsic ability of a system to relax towards equilibrium 
\cite{gyulassy},
especially for $\zeta=0$ which is, in particular, the case of CFT.

\section{Computing $\eta /s $ from the AdS/CFT correspondence}

How is it possible to compute the diffusion coefficients? The
traditional way is by using kinetic theory, more precisely the
Boltzmann equation or its quantum version, the
Uehling-Uhlenbenck equation, written in terms of the elastic
cross-sections, and  which can be solved by the
Chapman-Enskog method to first order in the perturbation out of 
equilibrium \cite{llanesdobado2003}. 
When  can we then apply these kinetic theory methods? 

 The main condition is that the mean free path must be much larger than the
interaction distance, which typically occurs for low density, weakly
interacting, systems. Then one can find the well known and somewhat
counter-intuitive result establishing that the greater the
interaction, the lesser the viscosity.

From a more modern prospective, it is possible to use the so
called Kubo formulae. The corresponding shear viscosity (for
short, viscosity in the following) can be obtained by analyzing linear
response theory and coupling the fluid to gravity by using an
appropriate lightly curved space-time background. Then one can
find:
\begin{equation}
\eta= \lim_{\omega\rightarrow 0}\left(\frac{1}
 {2\omega}\int dt d \bar x
e^{i\omega t}\langle[T_{xy}(t,\bar x),T_{xy}(0,\bar 0) ]\rangle 
\right) \ .
\end{equation}
This equation can be used to compute $\eta/s$ in the context of
the AdS/CFT correspondence. In order to do that one starts from a
CFT with gravity dual. For example for ${\mathcal{N}}=4, SU(N_c)$ Super
Yang-Mills one can consider the metric in five-dimensional AdS
\begin{equation}
ds^2=\frac{r^2}{R^2}[-(1-\frac{r_0^4}{r^4})dt^2+dx^2+dy^2+dz^2]
+\frac{R^2}{r^2(1-r_0^4/r^4)}dr^2 \ .
\end{equation}
The dual theory is a CFT at temperature  $T$ equal to the
Hawking temperature of the black-brane and Bekenstein entropy
$S=A/4 G_N$, where $A$ is the (hyper)area of the black-brane
horizon located at $r=r_0$. Now, following Klebanov 
\cite{Klebanov:1997kc}, 
consider a graviton polarized in the x-y direction and propagating
perpendiculary to the brane. The absorption
cross-section of the graviton by the brane measures, in the dual CFT, 
the imaginary part of the retarded Green's function of the operator 
coupled to the
metric i.e. the energy-momentum tensor. Then it is possible to
find:
\begin{equation}
\eta=\frac{\sigma(0)}{16\pi G_N}
\end{equation}
where $\sigma(0)$ is the graviton absorption cross-section at zero
energy. This cross-section can be computed classically by using
linerized Einstein equations and it turns out equal to the
horizon area so that one finally arrives to the formula shown in
Eq.(\ref{thebound}). 
Quite remarkable is that the result then does not depend on the 
particular form of the metric. 
It is the same for Dp, M2 and M5 branes.
Basically the reason is the universality of the graviton
absorption cross-section.

\section{ The conjecture of Kovtun, Son and Starinets}
From the result for the $\eta/s$ ratio for theories with a gravity
dual, Kovtun, Son and Starinets (KSS) proposed the conjecture that, for
a very wide class of systems, including those that can be
described by a sensible (i.e. ultraviolet complete) quantum field theory,
 the above ratio has the lower bound:
\begin{equation}
\frac{\eta}{s} \geq \frac{1}{4\pi}.
\end{equation}
There are several pieces of evidence supporting this conjecture.
The first one is based on the Heisenbeg uncertainty principle and
kinetic theory. The viscosity is proportional to the energy
density $\epsilon$ and the mean free time $\tau$. On the other
hand the entropy density is proportional to the number density
$n$. Therefore $\eta/s \sim \epsilon \tau/n \sim E \tau $ where
$E$ is the average particle energy. Thus, from the Heisenberg
principle for time-energy, we obtain the bound modulo the
numerical constant. 

Another important hint to establish the bound
comes from the computation done by Buchel, Liu and Starinets
\cite{buchel} where they obtained the leading correction to the
$\eta/s$ parameter in inverse powers of the 't Hooft coupling
using the $\alpha'$ corrected low energy effective action for the
typeIIB string theory, dual to the ${\mathcal{N}}=4$, $SU(N_c)$ SYM. 
They found the beautiful  result:
\begin{equation}
\frac{\eta}{s}=\frac{1}{4\pi}[1
+\frac{135\zeta(3)}{8(2g^2N_c)^{3/2}}+...]
\end{equation}
where $\zeta (z)$ is the Riemann $\zeta$-function so that
$\zeta(3)=1.2020569...$ is the Ap\'ery constant. Therefore the
correction to the ratio $\eta/s$ is positive and diverges for the
't Hooft coupling $\lambda$ approaching zero, and saturates the bound
for $\lambda$ growing to infinity. Assuming a smooth extrapolation
between the two regimes one can infer the validity of the bound
for any value of $\lambda$. Finally, we have knowledge of no fluid
that undercomes the bound. For all fluids examined so far, 
the $\eta/s$ ratio is well above the bound for the range of measured 
temperatures and preasures. This includes also superfluid helium and 
even trapped $ ^6{\rm Li}$ at strong coupling \cite{schaefer}.

An obvious consequence of the KSS conjecture (if correct), 
is the absence of perfect fluids in Nature, since the entropy density
will only vanish at absolute zero. Is this physically acceptable?
Of course, ideal fluids have been a source of interesting paradoxes.

Already in non relativistic fluid dynamics one encounters 
the d'Alembert paradox (an ideal fluid with no boundaries exerts no 
force on a body moving through it, i.e. there is no lift force). 
From it follows the umconfortable impossibilityh of flying or 
swimming. Of course, the difficulty disappears if there are no perfect 
fluids.

As a more recent conceptual difficulty of ideal fluids, 
we recall that it has been pointed out that the accretion of an
ideal fluid onto a black hole could violate the Generalized Second
Law of  Thermodynamics \cite{bekenstein}, suggesting a possible
connection between this law and the KSS bound.

In conclusion, the existence of a minimum viscosity would put to rest a 
number of problems in fluid mechanics, so the  
bound is not unwelcome by theorists.

\section {The RHIC case}
The 3834 m long Relativistic Heavy Ion Collider (RHIC) is operated at 
the Brookhaven National Laboratory (BNL) for, among others,
Au+Au collisions (A= 197), and can reach a center of mass energy
per nucleon of $E=200$ GeV with a luminosity  ${\mathcal{L}}= 2\times 
10^{26}
cm^{-2} s^{-1}$. It has four experiments called STAR, PHENIX,
BRAHMS and PHOBOS. We highlight some (preliminary) results from 
RHIC:
\begin{itemize}
\item
 Thermochemical models describe well the different particle yields 
fitting to $T=177 MeV$ and baryon chemical potential $\mu_B = 29 MeV$  
for  $E_{CM}  = 200 GeV$ per nucleon.
\item
From observed transversity and rapidity distributions, the Bjorken
model predicts an energy density at time $t_0 = 1 fm$  of $4 GeV
fm^{-3}$ whereas the critical density is about $0.7 GeV fm^{-3}$,
i.e. the matter created may be well above the threshold for Quark
Gluon Plasma (QGP) formation.
\item
A surprising amount of collective flow is observed in the outgoing
hadrons \cite{shuryak}, both in the single particle transverse
momentum distribution (radial flow) and in the asymmetric
azimuthal distribution around the beam axis (elliptic flow).
\end{itemize}
For our purposes, the main conclusions  derivable from these 
preliminary results are:
\begin{itemize}
\item Fluid dynamics with very low viscosity reproduces the measurements
of radial and elliptic flow up to transverse momenta of 1.5 GeV.

\item Collective flow is probably generated early in the collision,
probably in the QGP phase before hadronization. The QGP seems to
be more strongly interacting than expected on the basis of
perturbative QCD and asymptotic freedom (hence the low viscosity). An
alternative exists,  fast, instability-driven, equilibration
\cite{Mrowczyski}.

\item Some preliminary estimations of $\eta/s$ based on elliptic flow
\cite{shuryak,Teaney} and transverse momentum correlations
\cite{gavin} seem to be compatible with  value close to 0.08 (the
KSS bound). This would make it the most perfect fluid known.
\end{itemize}
\section{ Can the KSS bound be violated?}
Several avenues to theoretically break the KSS bound have been 
pursued. If one tries to undercome a bound on $\eta/s$, the 
possible strategies are to either decrease the viscosity at fixed 
entropy density, or to increase the entropy density at fixed viscosity.
We now briefly comment on these attempts and their status.

\subsection{Increase the entropy mixing several species}

 This possibility was realized already in \cite{son3} and has been
exploited in detail in \cite{cohen} not without some controversy
\cite{son4}. The  basic idea is that in principle it could be
possible to avoid the KSS bound in a non-relativistic  system with
constant cross section and a large number $g$ of non-identical,
degenerate particles, by increasing the Gibbs mixing entropy.

It is possible that the KSS bound  applies only to systems 
that can be obtained from a ``sensible'' (UV complete) QFT. Is it 
possible to find a non-relativistic system coming from a sensible QFT 
that violates the KSS bound for large degeneracy? As an interesting
example we can consider the case of a massive Non Linear Sigma
Model (NLSM) based on the coset $SO(g+1)/SO(g)$. Then in the non
relativistic limit it is possible to find \cite{Dobado:2007tm}:
\begin{equation}
\frac{\eta}{s}=\frac{80\sqrt{2}\pi^3}{11}\frac{f^4}{m^4}\frac{m}{T}
\frac{1}{n\lambda^3(\log\frac{g}{n\lambda^3}+\frac{5}{2})}
\end{equation}
where $m$ is the mass of the $g$ degenerated pseudo Goldstone
bosons, $f$ is the NLSM scale parameter, $n$ is the particle
number density and $\lambda$ is the thermal de Broglie wavelength
$\lambda=\sqrt{2\pi/m T}$. The above result is valid only in the
region $T<<m, n\lambda ^3<<g$ and $m \sim f$. Inside this region
the KSS bound is satisfied provided $g$ is not very large, but it
fails  if $g$ is exponentially large.

The NLSM is not an
UV complete theory but in principle we have at least two ways to
complete it.

First, simply QCD, since the above NLSM is the lagrangian of
Chiral Perturbation Theory at the lowest order with $g=3$ for
$N_f=2$ degenerated quarks since $SU(2)\times SU(2)/ SU(2) =
SO(4)/SO(3)$. This corresponds to having three Goldstone bosons
(pions). Then it would appear that increasing the number of the
QCD flavors, which implies increasing the number of Goldstone
bosons, one could undermine the KSS bound according to the formula
above. However this is not the case since, at it is well known,
one cannot increase the QCD flavor number without changing the
sign the derivative of the $\beta$ function at the origin,
$\beta(g)=-g^3(11-2N_f/3)/16\pi^2+..$ (here $g$ is the QCD
coupling, not the pion degeneracy) and presumably QCD is not well
defined outside the asymptotic freedom regime. In addition the
above formula applies only for NLSM coset $SO(g+1)/SO(g)$ whilst
the chiral QCD coset is $SU(N_f)_L\times SU(N_f)_R/
SU(N_f)_{L+R}$. Both families of cosets meet only for $g=3$ and
$N_f=2$ but not in the general case. Thus the above results does
not describe low-energy QCD for $g$ different from 3.

The other possibility for completing the NLSM is by introducing
the corresponding Liner Sigma Model (LSM) since the NLSM can be
considered as the low energy limit of the LSM for large $\sigma$
(Higgs) masses and the LSM is, at least perturbatively a well
defined renormalizable QFT. In this case, and for the same regime of
validity than in the NLSM, we get the same result for $\eta/s$.
Thus, for an exponentially large $g$ we can  violate the KSS
bound. Is there any way out for the KSS conjecture? In
principle there is one. The above computation only considered binary
interactions, but even at very small temperatures we have an
exponential number of non-elastic processes which make the
system metastable with a very weak tendency to relax to a state
where the density is fixed by the condition $\mu=m$ and where the
result above does not apply any more. Another possibility to
keep the validity of the KSS bound is the almost established
triviality of the LSM which would ruin this way to complete the
NLSM through a {\it sensible} QFT theory.

\begin{figure}
\includegraphics[height=2.0in]{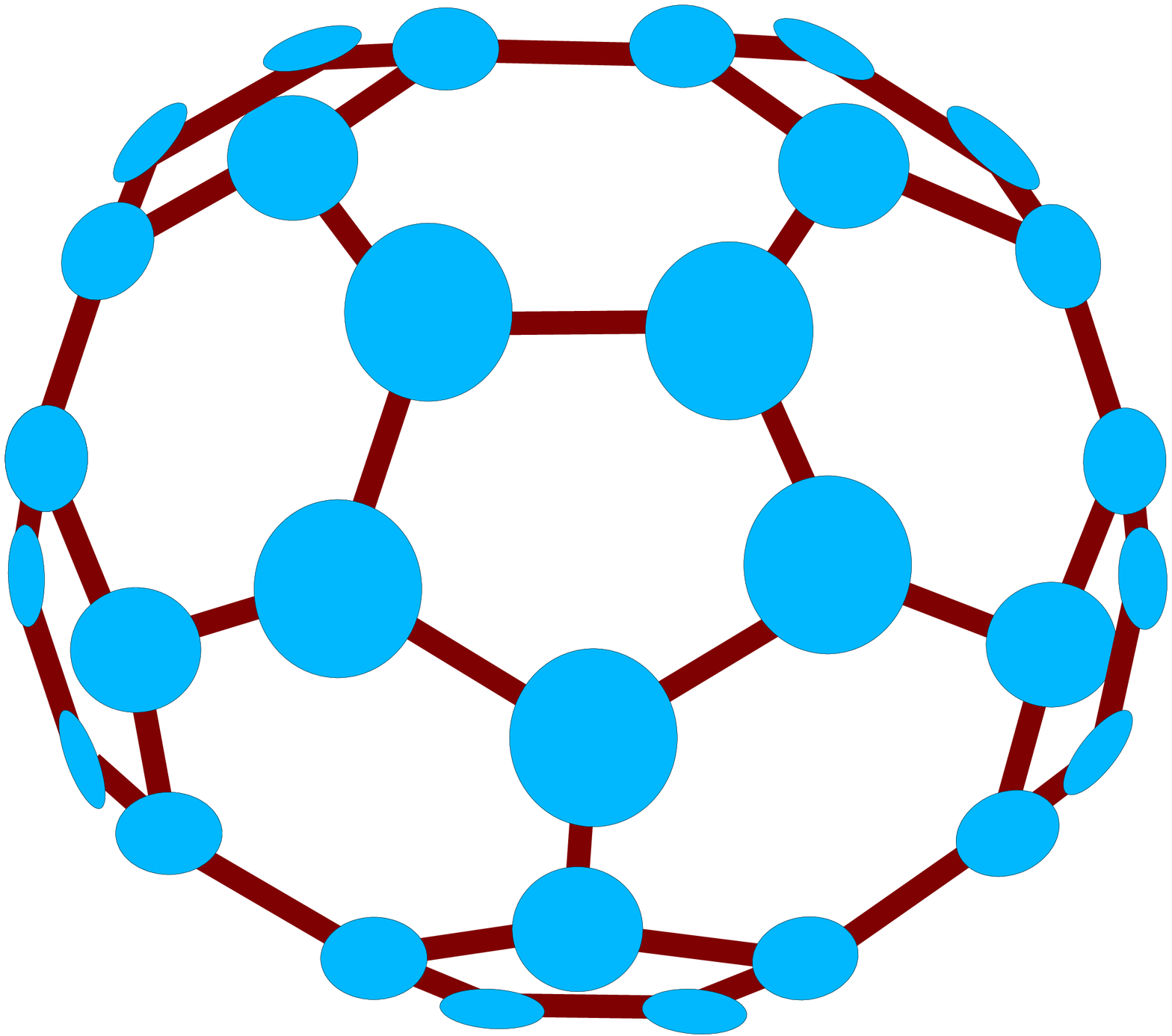}
\includegraphics[height=2.0in]{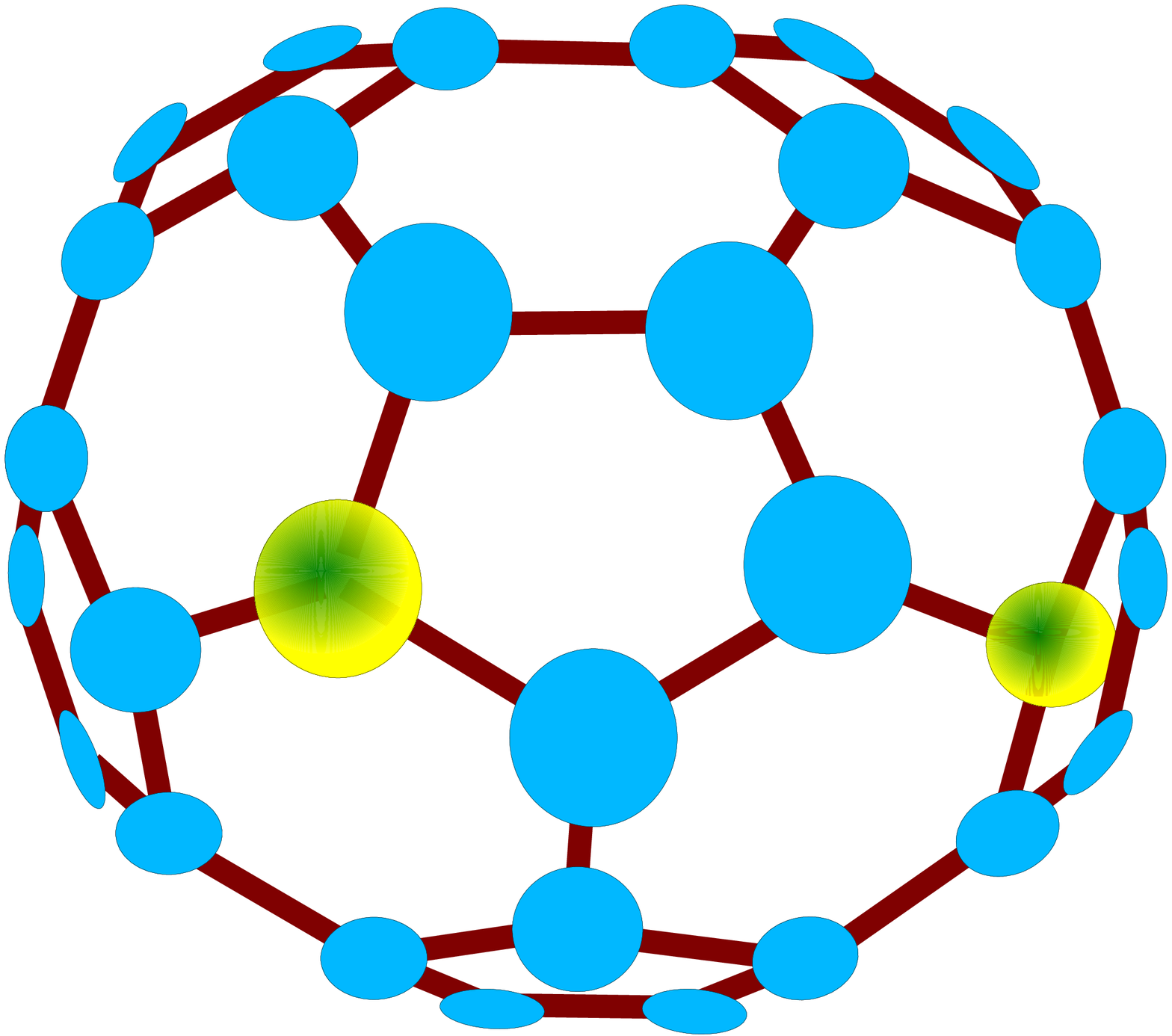}
\caption{A way to violate the KSS bound is by employing the logarithmic
increase of the mixing entropy in a multicomponent gas. This might be
achieved with a factorially large number of molecules, here a cartoon of
fullerene with two substitutions. However, simple estimates
\cite{Dobado:2007tm} indicate that it will be difficult to achieve fast
enough growth of the number of species with the molecule size (since
the sublimation temperature and hence viscosity also grow with the
size).
\label{fig:ful}}
\end{figure}

The concept of multiple substitutions in complex molecules, yielding an 
exponentially (factorially) large number of isomers, is illustrated in 
figure \ref{fig:ful}. Although we have not been able to clearly identify 
a family of currently synthesyzed  molecules, because of the complicated 
interplay of tightness of binding, excluded volume, sublimation 
temperature, etc. that vary together with the number of substitutable 
nodes, the possibility remains a priori. 
It is not clear how the argument of ultraviolet completion and a 
discussion about the completion of the standard model would affect 
a non-relativistic gas (the relevant case for most of our physical 
reality). From a practical point of view, although we know of no fluid 
that violates the bound, it is not obvious how this empirical fact is 
tied to the ``weak'' form of the conjecture that requires good UV 
behavior, since by arguments of scale separation, the low-energy 
effective theory would be all needed to describe reality to a given 
precision. Investigations continue.

\subsection{Decrease the viscosity by altering the dual gravitational 
action}
Another possible source of violation of the bound comes from
considering changes of the scattering cross-sections in the gas that
decrease the viscosity. 

From the point of view of the AdS/CFT
correspondence, the bound is saturated for standard general
relativity. However, on general grounds, one expects to have
higher derivative corrections to the Einstein-Hilbert action. For
example one could consider the case of a theory with a gravity
dual described at low energy by a Gauss-Bonnet gravity action:
\begin{equation}
S_{GB}=\frac{1}{16\pi G_N}=\int
dx^5\sqrt{-g}[R-2\Lambda+\frac{\lambda_{GB}}{2}L^2(R^2-4R_{\mu\nu}
R^{\mu\nu}+R_{\mu\nu\varrho\sigma} R^{\mu\nu\varrho\sigma})]
\end{equation}
where $\Lambda=-6/L^2$. In this case it is possible to find
\footnote{The authors thank Juan Maldacena for bringing our
attention to these references.} \cite{corrections}:
\begin{equation}
\frac{\eta}{s}=\frac{1}{4\pi}(1-4\lambda_{GB})
\end{equation}
This result is non perturbative in $\lambda_{GB}$ and shows that
the bound can be violated for positive $\lambda_{GB}$. One could argue
that this result is suspicious since it does not make sense for
$\lambda_{GB} > 4$. However in \cite{causality} it is shown that,
in order to avoid microcausality violations in the corresponding
CFT we need to have
\begin{equation}
\frac{\eta}{s}\geq \frac{16}{25}(\frac{1}{4\pi}).
\end{equation}
This leaves room for violation of the KSS bound but requires
$\lambda_{GB}\leq 9/100$ for the Gauss-Bonnet theory to  be
consistent. However there could be other additional  consistency
constraints in the theory forcing the $\lambda_{GB}$ parameter to
vanish or be negative thus reestablishing the validity of the
KSS bound.

\subsection{Unitarity invalidates certain reported violations}
\begin{figure}[h]
\includegraphics[height=2.8in]{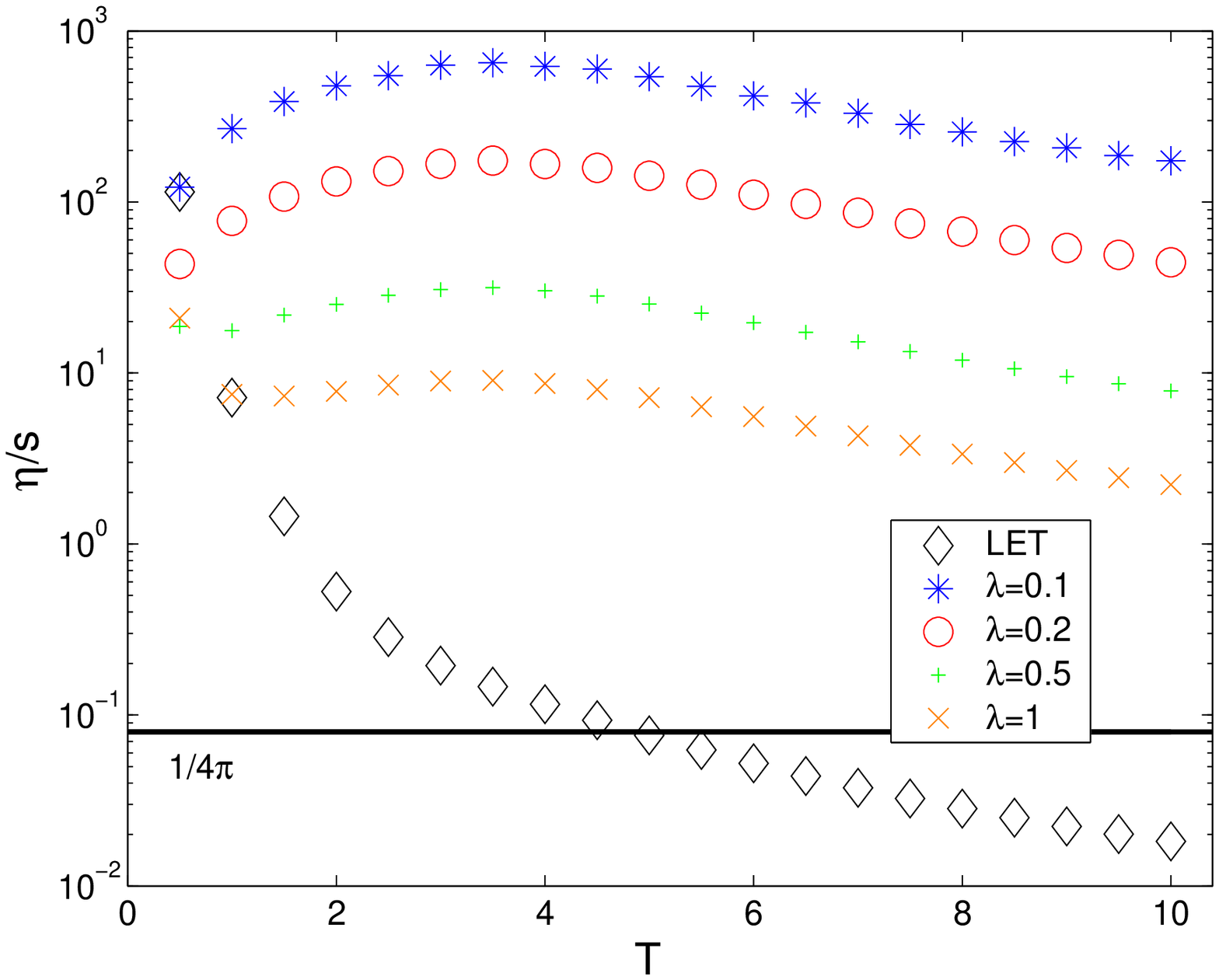}
\includegraphics[height=2.8in]{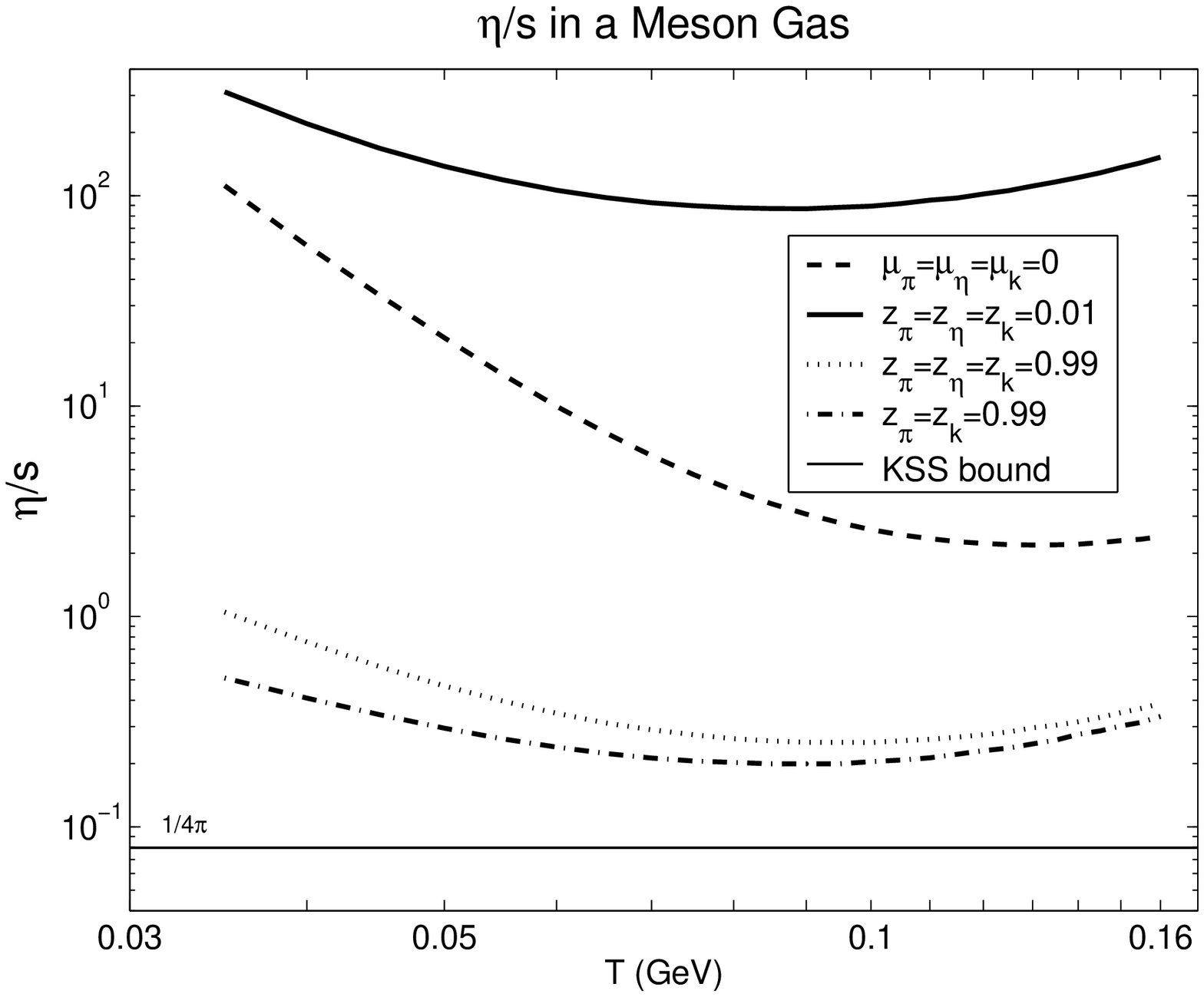}
\caption{Reported violations of the $\eta/s$ holographic bound in the
meson gas at high temperature are spurious: in this graph we show how
the cross section computed within unitarized Chiral Perturbation Theory
by means of the Inverse Amplitude Method, respects the bound $1/4\pi$
for all conceivable temperatures that the hadron gas might attain
(right). Violations reported are due to an unphysically small (large)
viscosity (cross section) induced by unitarity violations.
For comparison, we show a similar calculation with the
unitarized Linear Sigma Model (left) that yields the same result.
\label{fig:unitarity}}
\end{figure}

From the QFT point of view, a way to decrease the viscosity without 
affecting the entropy in excess is to increase the interparticle 
interaction. This has been remarked in the past, in particular in the 
context of the pion gas. We however have shown 
that the method fails because unitarity imposes a bound on the 
cross-section.

Thus, the extrapolation of the low-energy effective theory for pions 
(ChPT to first or second order) to higher energies, ceases to be valid 
when resonances in the meson gas are reached ($\sigma$, $\rho$, etc.). A 
properly unitarized method such as the IAM has been reported in 
reference \cite{Dobado:2006hw} and is plotted in figure 
\ref{fig:unitarity}, where we have extended our past calculations to 
include strange mesons ($K$, $\eta$) and their lowest resonances 
through the $SU(3)$ IAM \cite{Pelaez:2001db}.

\section{$\eta/s$ and the phase transition}
Recently Csernai, Kapusta and McLerran \cite{CKM} made the
observation that, in all systems whose $\frac{\eta}{s}(T)$ plot 
has been examined, the minimum of $\eta/s$ and
the liquid-gas phase transition happen at the same temperature. 
This  fact can be roughly understood because in a gas, as the 
temperature increases, there is more efficient momentum transport and 
then the viscosity goes to infinity.
However in a liquid, which can be seen as a mixture of clusters
and voids, molecules swap voids less efficiently at low 
temperature.
Then as temperature decreases there are less voids and
consequently more shear momentum transport due to adhesivity among 
layers, so that the viscosity takes also off to infinity. 

Therefore somewhere between the two phases the
viscosity should have a minimum. Empirically, for standard fluids,
$\eta/s$  is found to reach its minimum at or near the critical
temperature. Thus, apparently there is a connection between
$\eta/s$ and the phase transition but we do not have any clear theory
of this \cite{Dobado:2008vt}. For example we ignore if there are 
universal critical exponents for the different kind of phase 
transitions. A very interesting possibility is that the same behavior 
observed in ordinary liquids could also occur in QCD. If this were the 
case, an experimental determination of the $\eta/s$ minimum could give
information about the QCD phase transition and whether there is a 
critical end-point. This direction has been explored in \cite{etapt}. 

\begin{figure}
\includegraphics[height=3.0in]{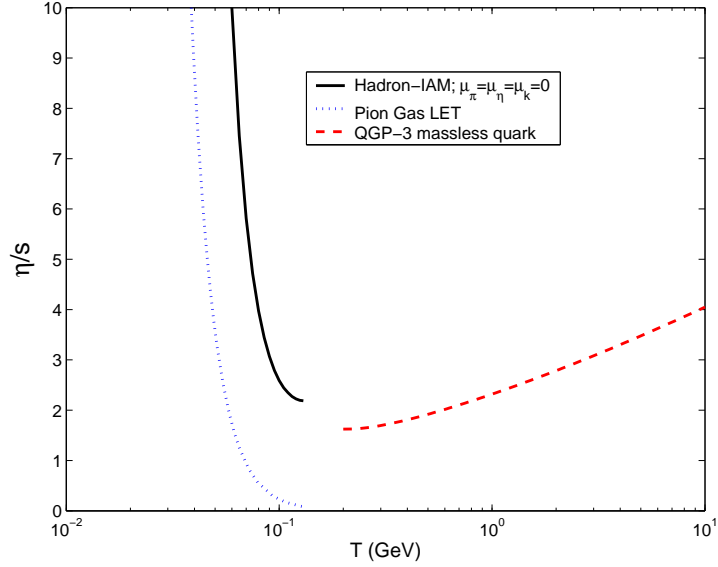}
\caption{ By analogy with common, human-scale fluids, $\eta/s$ is 
expected to have a minimum but not a discontinuity at the predicted 
cross-over between the hadron phase and the quark and gluon plasma 
phase. We have reexamined the computation in the hadron phase with more 
detailed than past work, but a discontinuity remains, maybe due to the 
uncertainties in the quark-gluon plasma side. This observable is a 
promising tag of a possible critical end point in the QCD phase-diagram.
\label{fig:salto}}
\end{figure}
The hadron-gas phase has been examined with great care by us  
\cite{Dobado:2008vt} and close colleagues 
\cite{FernandezFraile:2007zz,FernandezFraile:2007gd}, 
so that great improvements are not to be expected. The quark and gluon 
plasma phase, however, is less known in the strong-coupling regime and 
requires more work. The situation is represented in figure 
\ref{fig:salto}. At high baryon chemical potential one would expect a 
jump in the ratio $\eta/s$ (by analogy with common substances) at a 
first order phase transition, but a continuous minimum at low 
baryon-density where one expects a cross-over. Current computations of 
$\eta/s$ from the hadron phase are quite mature, but the quark-gluon 
plasma has not yet been refined beyond high-$T$ perturbation theory.

\section{ Conclusions and open questions}
The AdS/CFT correspondence makes possible to study new aspects of
QFT such as viscosity and other hydrodynamic behavior. The KSS
bound set a new limit on how perfect a fluid can be coming from
holography which was completely unexpected. From the experimental
stand point there is no counter example for this bound. From the RHIC
data we observe a large  amount of  collective flow that can be
properly described by hydrodynamic models with low viscosity
compatible with the KSS bound. Some theoretical models suggest
that unitarity could be related in some way with the KSS bound.
There is a theoretical counter example of the bound in a
non-relativistic model with large degeneracy. However possibly the
model is not UV complete because of the triviality of the LSM.
This could be an indication that a more precise formulation of the
bound is needed.

Some open questions are the following: Is the
bound correct for some well defined formulation? Could it be
possible to really measure  $\eta/s$ at RHIC with precision enough
to check the KSS bound? Are there any connections between the KSS
and the entropy Bekenstein bounds? How are the
minima of $\eta/s$ related to phase transitions? Could it be considered
an order parameter?

\vskip 1.0cm

 {\bf Acknowledgements}
 This work has been partially supported by the DGICYT (Spain) under 
 grants FPA 2004-02602 and FPA 2005-02327 and by the Universidad
 Complutense/CAM, project number 910309 and BSCH-PR34/07-15875. 
A. D. thanks Jos\'e Edelstein 
and the organization for their kind invitation to
 participate in this celebration of the tenth anniversary  of the
discovery of the AdS/CFT correspondence held at such a great place
as Buenos Aires, and Alex Buchel and Juan Maldacena for useful
comments.


\end{document}